\documentclass[11pt]{article}
\usepackage[a4paper]{geometry}
\usepackage{amsfonts, amsmath, amssymb, amsthm, graphicx, caption, authblk, multirow, makecell, framed, float, xcolor, enumitem, tikz, hyperref}
\theoremstyle{definition}
\newtheorem{theorem}{Theorem}

\theoremstyle{remark}

\newcommand*{\mybox}[1]{%
  \framebox{\raisebox{0cm}[0.5\baselineskip][0.05\baselineskip]{%
    \hbox to 0.10cm {\hss#1\hss}}}\hspace{0.05cm}}

\begin{document}
\title{Sumplete is Hard, Even with Two Different Numbers}
\author[1]{Suthee Ruangwises\thanks{\texttt{ruangwises@gmail.com}}}
\affil[1]{Department of Informatics, The University of Electro-Communications, Tokyo, Japan}
\date{}
\maketitle

\begin{abstract}
Sumplete is a logic puzzle famous for being developed by ChatGPT. The puzzle consists of a rectangular grid, with each cell containing a number. The player has to cross out some numbers such that the sum of uncrossed numbers in each row and column is equal to a given integer assigned to that row or column. In this paper, we prove that deciding solvability of a given Sumplete puzzle is NP-complete, even if the grid contains only two different numbers.

\end{abstract}

\section{Introduction}
\textit{Sumplete} is a logic puzzle similar to Sudoku, Kakuro, and Hitori. It is famous for being developed with the help of ChatGPT \cite{about}. The Sumplete puzzle consists of a rectangular grid, with each cell containing a single-digit positive number. Each row and column also has a non-negative integer called \textit{hint} assigned to it. The player has to cross out some numbers in the grid such that the sum of uncrossed numbers in each row and column is equal to the corresponding hint \cite{game}. See Figure \ref{fig1}.

\begin{figure}[H]
\centering
\begin{tikzpicture}
\draw[step=0.8cm,color=black] (0,0) grid (4,4);

\node at (0.4,-0.4) {11};
\node at (1.2,-0.4) {18};
\node at (2.0,-0.4) {11};
\node at (2.8,-0.4) {9};
\node at (3.6,-0.4) {10};

\node at (4.4,0.4) {15};
\node at (4.4,1.2) {6};
\node at (4.4,2.0) {11};
\node at (4.4,2.8) {14};
\node at (4.4,3.6) {13};

\node at (0.4,0.4) {3};
\node at (1.2,0.4) {3};
\node at (2.0,0.4) {4};
\node at (2.8,0.4) {9};
\node at (3.6,0.4) {6};
\node at (0.4,1.2) {6};
\node at (1.2,1.2) {2};
\node at (2.0,1.2) {4};
\node at (2.8,1.2) {9};
\node at (3.6,1.2) {4};
\node at (0.4,2.0) {4};
\node at (1.2,2.0) {7};
\node at (2.0,2.0) {2};
\node at (2.8,2.0) {5};
\node at (3.6,2.0) {2};
\node at (0.4,2.8) {5};
\node at (1.2,2.8) {1};
\node at (2.0,2.8) {4};
\node at (2.8,2.8) {1};
\node at (3.6,2.8) {8};
\node at (0.4,3.6) {3};
\node at (1.2,3.6) {5};
\node at (2.0,3.6) {5};
\node at (2.8,3.6) {7};
\node at (3.6,3.6) {1};
\end{tikzpicture}
\hspace{1.5cm}
\begin{tikzpicture}
\draw[step=0.8cm,color=black] (0,0) grid (4,4);

\node at (0.4,-0.4) {11};
\node at (1.2,-0.4) {18};
\node at (2.0,-0.4) {11};
\node at (2.8,-0.4) {9};
\node at (3.6,-0.4) {10};

\node at (4.4,0.4) {15};
\node at (4.4,1.2) {6};
\node at (4.4,2.0) {11};
\node at (4.4,2.8) {14};
\node at (4.4,3.6) {13};

\node at (0.4,0.4) {3};
\node at (1.2,0.4) {3};
\node [opacity=0.5] at (2.0,0.4) {\scalebox{4}{$\times$}};
\node at (2.0,0.4) {4};
\node at (2.8,0.4) {9};
\node [opacity=0.5] at (3.6,0.4) {\scalebox{4}{$\times$}};
\node at (3.6,0.4) {6};
\node [opacity=0.5] at (0.4,1.2) {\scalebox{4}{$\times$}};
\node at (0.4,1.2) {6};
\node at (1.2,1.2) {2};
\node at (2.0,1.2) {4};
\node [opacity=0.5] at (2.8,1.2) {\scalebox{4}{$\times$}};
\node at (2.8,1.2) {9};
\node [opacity=0.5] at (3.6,1.2) {\scalebox{4}{$\times$}};
\node at (3.6,1.2) {4};
\node [opacity=0.5] at (0.4,2.0) {\scalebox{4}{$\times$}};
\node at (0.4,2.0) {4};
\node at (1.2,2.0) {7};
\node at (2.0,2.0) {2};
\node [opacity=0.5] at (2.8,2.0) {\scalebox{4}{$\times$}};
\node at (2.8,2.0) {5};
\node at (3.6,2.0) {2};
\node at (0.4,2.8) {5};
\node at (1.2,2.8) {1};
\node [opacity=0.5] at (2.0,2.8) {\scalebox{4}{$\times$}};
\node at (2.0,2.8) {4};
\node [opacity=0.5] at (2.8,2.8) {\scalebox{4}{$\times$}};
\node at (2.8,2.8) {1};
\node at (3.6,2.8) {8};
\node at (0.4,3.6) {3};
\node at (1.2,3.6) {5};
\node at (2.0,3.6) {5};
\node [opacity=0.5] at (2.8,3.6) {\scalebox{4}{$\times$}};
\node at (2.8,3.6) {7};
\node [opacity=0.5] at (3.6,3.6) {\scalebox{4}{$\times$}};
\node at (3.6,3.6) {1};
\end{tikzpicture}
\caption{An example of a $5 \times 5$ Sumplete puzzle (left) and its solution (right)}
\label{fig1}
\end{figure}

\subsection{Our Contribution}
In this paper, we prove that deciding whether a given Sumplete instance has a solution is NP-complete, even if the grid contains only two different numbers. In fact, we prove for the case where the grid contains only 1s and 3s. We call such instance a \textit{$(1,3)$-Sumplete instance}.

\begin{theorem}
Deciding solvability of a given $(1,3)$-Sumplete instance is NP-complete.
\end{theorem}

As the problem clearly belongs to NP, the more challenging part is to prove the NP-hardness. We do so by constructing a reduction from the XSAT problem for $\text{3-CNF}_{+}^{3}$ (deciding if there is a Boolean assignment such that every clause has exactly one literal that evaluates to true, in a setting where each clause contains exactly three positive literals and each variable appears in exactly three clauses), which is known to be NP-complete \cite{sat}.

\section{Reduction}
Suppose we have $n$ variables $x_1,x_2,...,x_n$ and $n$ clauses $c_1,c_2,...,c_n$ (for $\text{3-CNF}_{+}^{3}$, the number of variables and number of clauses must be equal). We construct a $(1,3)$-Sumplete instance with size $(n+1) \times n$ as follows.

Let $a(i,j)$ denote the number in the $i$-th row and $j$-th column. For $i \leq n$, we set
$$a(i,j) = \begin{cases}
  1, &x_j \text{ appears in } c_i \\
  3, &x_j \text{ does not appear in } c_i.
\end{cases}$$
Also, we set $a(n+1,j)=3$ for every $j=1,2,...,n$.

Let $R(i)$ denote the hint of the $i$-th row and $C(j)$ denote the hint of the $j$-th column. We set $R(i)=1$ for $i=1,2,...,n$ and $R(n+1)=2n$. Also, we set $C(j)=3$ for every $j=1,2,...,n$. See Figures \ref{fig2} and \ref{fig3}.

\subsection{Proof of Correctness}
Let literal $x_j$ in clause $c_i$ evaluate to true if and only if $a(i,j)$ is uncrossed.

Due to the hints of the first $n$ rows, there can be only one uncrossed number in each of the first $n$ rows, and it must be a 1. This is equivalent to the condition that exactly one literal in each clause is true.

Due to the hints of the columns, the sum of uncrossed numbers in the first $n$ rows of each column must be either 0 or 3, which means either all numbers are crossed or only the three 1s are uncrossed (since all 3s in the first $n$ rows must be crossed). This is equivalent to the condition that a variable must be true in all clauses it appears, or false in all clauses it appears.

Therefore, the Sumplete instance has a solution if and only if the original XSAT problem is satisfiable. Note that the hint of the $(n+1)$-th row always holds since the sum of all uncrossed numbers in the grid is $3n$.

\begin{figure}[H]
\centering
\begin{tikzpicture}
\node at (0,0) {$c_1 = x_1 \vee x_2 \vee x_3$};
\node at (0,-0.5) {$c_2 = x_2 \vee x_3 \vee x_6$};
\node at (0,-1) {$c_3 = x_1 \vee x_4 \vee x_6$};
\node at (0,-1.5) {$c_4 = x_2 \vee x_5 \vee x_6$};
\node at (0,-2) {$c_5 = x_1 \vee x_4 \vee x_5$};
\node at (0,-2.5) {$c_6 = x_3 \vee x_4 \vee x_5$};
\end{tikzpicture}
\hspace{4cm}
\begin{tikzpicture}
\node at (0,0) {$x_1 = \text{FALSE}$};
\node at (0,-0.5) {$x_2 = \text{TRUE}$};
\node at (0,-1) {$x_3 = \text{FALSE}$};
\node at (0,-1.5) {$x_4 = \text{TRUE}$};
\node at (0,-2) {$x_5 = \text{FALSE}$};
\node at (0,-2.5) {$x_6 = \text{FALSE}$};
\end{tikzpicture}
\caption{An XSAT problem for $\text{3-CNF}_{+}^{3}$ (left) and its solution (right)}
\label{fig2}
\end{figure}

\begin{figure}[H]
\centering
\begin{tikzpicture}
\draw[step=0.8cm,color=black] (0,0) grid (4.8,5.6);

\node at (0.4,-0.4) {3};
\node at (1.2,-0.4) {3};
\node at (2.0,-0.4) {3};
\node at (2.8,-0.4) {3};
\node at (3.6,-0.4) {3};
\node at (4.4,-0.4) {3};

\node at (5.2,0.4) {12};
\node at (5.2,1.2) {1};
\node at (5.2,2.0) {1};
\node at (5.2,2.8) {1};
\node at (5.2,3.6) {1};
\node at (5.2,4.4) {1};
\node at (5.2,5.2) {1};

\node at (0.4,0.4) {3};
\node at (1.2,0.4) {3};
\node at (2.0,0.4) {3};
\node at (2.8,0.4) {3};
\node at (3.6,0.4) {3};
\node at (4.4,0.4) {3};
\node at (0.4,1.2) {3};
\node at (1.2,1.2) {3};
\node at (2.0,1.2) {1};
\node at (2.8,1.2) {1};
\node at (3.6,1.2) {1};
\node at (4.4,1.2) {3};
\node at (0.4,2.0) {1};
\node at (1.2,2.0) {3};
\node at (2.0,2.0) {3};
\node at (2.8,2.0) {1};
\node at (3.6,2.0) {1};
\node at (4.4,2.0) {3};
\node at (0.4,2.8) {3};
\node at (1.2,2.8) {1};
\node at (2.0,2.8) {3};
\node at (2.8,2.8) {3};
\node at (3.6,2.8) {1};
\node at (4.4,2.8) {1};
\node at (0.4,3.6) {1};
\node at (1.2,3.6) {3};
\node at (2.0,3.6) {3};
\node at (2.8,3.6) {1};
\node at (3.6,3.6) {3};
\node at (4.4,3.6) {1};
\node at (0.4,4.4) {3};
\node at (1.2,4.4) {1};
\node at (2.0,4.4) {1};
\node at (2.8,4.4) {3};
\node at (3.6,4.4) {3};
\node at (4.4,4.4) {1};
\node at (0.4,5.2) {1};
\node at (1.2,5.2) {1};
\node at (2.0,5.2) {1};
\node at (2.8,5.2) {3};
\node at (3.6,5.2) {3};
\node at (4.4,5.2) {3};
\end{tikzpicture}
\hspace{1.5cm}
\begin{tikzpicture}
\draw[step=0.8cm,color=black] (0,0) grid (4.8,5.6);

\node at (0.4,-0.4) {3};
\node at (1.2,-0.4) {3};
\node at (2.0,-0.4) {3};
\node at (2.8,-0.4) {3};
\node at (3.6,-0.4) {3};
\node at (4.4,-0.4) {3};

\node at (5.2,0.4) {12};
\node at (5.2,1.2) {1};
\node at (5.2,2.0) {1};
\node at (5.2,2.8) {1};
\node at (5.2,3.6) {1};
\node at (5.2,4.4) {1};
\node at (5.2,5.2) {1};

\node at (0.4,0.4) {3};
\node at (1.2,0.4) {3};
\node at (2.0,0.4) {3};
\node at (2.8,0.4) {3};
\node at (3.6,0.4) {3};
\node at (4.4,0.4) {3};
\node at (0.4,1.2) {3};
\node at (1.2,1.2) {3};
\node at (2.0,1.2) {1};
\node at (2.8,1.2) {1};
\node at (3.6,1.2) {1};
\node at (4.4,1.2) {3};
\node at (0.4,2.0) {1};
\node at (1.2,2.0) {3};
\node at (2.0,2.0) {3};
\node at (2.8,2.0) {1};
\node at (3.6,2.0) {1};
\node at (4.4,2.0) {3};
\node at (0.4,2.8) {3};
\node at (1.2,2.8) {1};
\node at (2.0,2.8) {3};
\node at (2.8,2.8) {3};
\node at (3.6,2.8) {1};
\node at (4.4,2.8) {1};
\node at (0.4,3.6) {1};
\node at (1.2,3.6) {3};
\node at (2.0,3.6) {3};
\node at (2.8,3.6) {1};
\node at (3.6,3.6) {3};
\node at (4.4,3.6) {1};
\node at (0.4,4.4) {3};
\node at (1.2,4.4) {1};
\node at (2.0,4.4) {1};
\node at (2.8,4.4) {3};
\node at (3.6,4.4) {3};
\node at (4.4,4.4) {1};
\node at (0.4,5.2) {1};
\node at (1.2,5.2) {1};
\node at (2.0,5.2) {1};
\node at (2.8,5.2) {3};
\node at (3.6,5.2) {3};
\node at (4.4,5.2) {3};

\node [opacity=0.5] at (1.2,0.4) {\scalebox{4}{$\times$}};
\node [opacity=0.5] at (2.8,0.4) {\scalebox{4}{$\times$}};
\node [opacity=0.5] at (0.4,1.2) {\scalebox{4}{$\times$}};
\node [opacity=0.5] at (1.2,1.2) {\scalebox{4}{$\times$}};
\node [opacity=0.5] at (2.0,1.2) {\scalebox{4}{$\times$}};
\node [opacity=0.5] at (3.6,1.2) {\scalebox{4}{$\times$}};
\node [opacity=0.5] at (4.4,1.2) {\scalebox{4}{$\times$}};
\node [opacity=0.5] at (0.4,2.0) {\scalebox{4}{$\times$}};
\node [opacity=0.5] at (1.2,2.0) {\scalebox{4}{$\times$}};
\node [opacity=0.5] at (2.0,2.0) {\scalebox{4}{$\times$}};
\node [opacity=0.5] at (3.6,2.0) {\scalebox{4}{$\times$}};
\node [opacity=0.5] at (4.4,2.0) {\scalebox{4}{$\times$}};
\node [opacity=0.5] at (0.4,2.8) {\scalebox{4}{$\times$}};
\node [opacity=0.5] at (2.0,2.8) {\scalebox{4}{$\times$}};
\node [opacity=0.5] at (2.8,2.8) {\scalebox{4}{$\times$}};
\node [opacity=0.5] at (3.6,2.8) {\scalebox{4}{$\times$}};
\node [opacity=0.5] at (4.4,2.8) {\scalebox{4}{$\times$}};
\node [opacity=0.5] at (0.4,3.6) {\scalebox{4}{$\times$}};
\node [opacity=0.5] at (1.2,3.6) {\scalebox{4}{$\times$}};
\node [opacity=0.5] at (2.0,3.6) {\scalebox{4}{$\times$}};
\node [opacity=0.5] at (3.6,3.6) {\scalebox{4}{$\times$}};
\node [opacity=0.5] at (4.4,3.6) {\scalebox{4}{$\times$}};
\node [opacity=0.5] at (0.4,4.4) {\scalebox{4}{$\times$}};
\node [opacity=0.5] at (2.0,4.4) {\scalebox{4}{$\times$}};
\node [opacity=0.5] at (2.8,4.4) {\scalebox{4}{$\times$}};
\node [opacity=0.5] at (3.6,4.4) {\scalebox{4}{$\times$}};
\node [opacity=0.5] at (4.4,4.4) {\scalebox{4}{$\times$}};
\node [opacity=0.5] at (0.4,5.2) {\scalebox{4}{$\times$}};
\node [opacity=0.5] at (2.0,5.2) {\scalebox{4}{$\times$}};
\node [opacity=0.5] at (2.8,5.2) {\scalebox{4}{$\times$}};
\node [opacity=0.5] at (3.6,5.2) {\scalebox{4}{$\times$}};
\node [opacity=0.5] at (4.4,5.2) {\scalebox{4}{$\times$}};
\end{tikzpicture}
\caption{A $(1,3)$-Sumplace instance transformed from the XSAT problem in Figure \ref{fig2} (left) and its solution (right)}
\label{fig3}
\end{figure}

\end{document}